\documentstyle[epsfig]{mn}
\begin{document}

\title
[Submillimetre-wave lenses and cosmology] 
{Submillimetre-wave gravitational lenses and cosmology} 
\author
[A. W. Blain]
{
A. W. Blain\\
Cavendish Laboratory, Madingley Road,
Cambridge, CB3  0HE.
}
\maketitle

\begin{abstract}
One of the most direct routes for investigating the geometry of the Universe
is provided by the numbers of strongly magnified
gravitationally lensed galaxies as compared with those that are either 
weakly magnified or de-magnified. In the submillimetre waveband the relative 
abundance of strongly lensed galaxies is expected to be larger as compared 
with the optical or radio wavebands, both in the field and in clusters of galaxies. 
The predicted numbers depend on the properties of the population of 
faint galaxies in the submillimetre waveband, which was formerly very uncertain; 
however, recent observations of lensing clusters have reduced this uncertainty 
significantly and confirm that a large sample of galaxy--galaxy lenses could be 
detected and investigated using forthcoming facilities, including the {\it FIRST} 
and {\it Planck Surveyor} space missions and a large ground-based 
millimetre/submillimetre-wave interferometer array (MIA). 
We discuss how this 
sample could be used to impose limits to the values of cosmological 
parameters and the total density and form of evolution of the mass distribution 
of bound structures, even in the absence of detailed lens modeling for individual 
members of the sample. The effects of 
different world models on the form of the magnification bias expected in sensitive 
submillimetre-wave observations of clusters are also discussed, because an MIA 
could resolve and investigate images in clusters in detail. 
\end{abstract}  

\begin{keywords}
galaxies: evolution -- cosmology: observations -- cosmology: theory -- 
gravitational lensing -- large-scale structure of universe -- 
radio continuum: galaxies
\end{keywords}

\section{Introduction}

The importance of the statistical properties of gravitational lenses for 
investigating the geometry of the Universe has been discussed by Gott, Park \& 
Lee (1989) and Fukugita et al. (1992). A particularly promising route to 
constraining the value of the cosmological constant $\Omega_\Lambda$ (Carroll, 
Press \& Turner 1992; Kochanek 1996) would be offered by a determination of the 
relative abundance of lensed and unlensed galaxies and quasars. Important 
developments are expected in this field when the next generation of 
submillimetre-wave telescopes becomes available, because the abundance of 
lensed images, formed by both intervening galaxies and clusters, is expected to 
be significantly larger in this waveband as compared with other wavebands 
(Blain 1996a,b, 1997b).

A large sample, of perhaps several hundred gravitational lenses (Blain 1997a,c), 
could be compiled by combining the results of large-area 
extragalactic surveys using the {\it Planck Surveyor} (Bersanelli et al. 1996) and 
{\it FIRST} (Beckwith et al. 1993) space-borne telescopes with
sub-arcsec-resolution imaging of the detected sources using a large 
ground-based millimetre/submillimetre-wave interferometer array 
(MIA; Brown 1996; Downes 1996). The probability of lensing, and hence the size
of the sample, depends on the world model and the normalisation and form of 
evolution of the mass distribution of lensing galaxies. About 100 fainter lenses 
could also be detected in a small-area survey using an MIA alone (Blain 1996a,b). 
Here the prospects for using such large samples of galaxy--galaxy lenses to 
investigate the values of cosmological parameters and the evolution of 
large-scale structures are discussed. We also consider whether the magnification 
bias of distant galaxies in the field of a cluster of galaxies (Broadhurst, Taylor \& 
Peacock 1995; Blain 1997b), which also depends on the world model, could be 
used to impose similar constraints. 

Existing observations in the optical waveband have allowed constraints to be 
imposed on the values of cosmological parameters. The observed numbers of 
lensed quasars appear to rule out world models with large values of
$\Omega_\Lambda$; Kochanek (1996) derives $\Omega_\Lambda < 0.66$ at a 
confidence level of 95\% in a flat world model. Observations of lensed images in 
clusters appear to favour a flat world model with a value of 
$\Omega_\Lambda \simeq 0.7$ (Fort, Mellier \& Dantel-Fort 1997; 
Wu \& Mao 1996).

The assumptions about world models and the form of evolution of distant 
galaxies in the present study are discussed in Section~2. The potential of 
submillimetre-wave observations of lensing by galaxies and clusters for 
investigating cosmology are discussed in Sections~3 and 4 respectively. 
In the sections of the paper that involve lensing by galaxies rather than clusters, 
galaxies that are multiply imaged and significantly magnified by lensing are 
described as `lensed', those that are not significantly magnified are described as 
`unlensed'. A value of Hubble's constant $H_0=50$\,km\,s$^{-1}$\,Mpc$^{-1}$ is 
assumed.

\section{The underlying models}

\begin{table}
\caption{A summary of the world models used in this paper. $t_0$ and 
$\Omega_{\rm m}$ represent the age of the Universe and the density parameter 
of metals (Section~2.3) respectively at the present epoch.}
\begin{tabular}{ p{1.1cm} p{1.1cm} p{1.2cm} p{1.5cm} p{1.4cm} } 
$\Omega_0$ & $\Omega_\Lambda$ & $t_0$/Gyr & Geometry & 
$\Omega_{\rm m}$/10$^{-4}$ \\
\noalign{\vskip 2mm}
1.0 & 0.0 & 13.1 & Flat & 4.5\\
0.7 & 0.3 & 14.7 & Flat & 5.2\\
0.3 & 0.7 & 18.9 & Flat & 7.4\\
0.16 & 0.84 & 22.3 & Flat & 9.4\\
0.7 & 0.0 & 14.0 & Open & 4.9\\
0.3 & 0.0 & 15.8 & Open & 5.9\\
0.16 & 0.0 & 16.9 & Open & 6.5\\
0.16 & 1.2 & 28.5 & Closed & 13\\
\end{tabular}
\end{table}

\subsection{Faint galaxies in the submillimetre waveband}

The form of the counts of galaxies in the submillimetre waveband is
uncertain at present. Predictions can be made by extrapolating the luminosity 
function of galaxies detected by the {\it IRAS} satellite at a wavelength of 
60\,$\mu$m (Saunders et al. 1990) out to large redshifts (Blain \& Longair 1993a); 
however, detailed knowledge will only be provided by the results of the first 
blank-field surveys in the submillimetre waveband (Blain \& Longair 1996). 
Evidence from observations in a range of wavebands indicates that the 
populations of star-forming and active galaxies can be adequately described by 
pure luminosity evolution of the local galaxy population with an approximate form 
$(1+z)^3$ out to a redshift $z \simeq 2$ (Dunlop \& Peacock 1990; Oliver, 
Rowan-Robinson \& Saunders 1992; Hewett, Foltz \& Chaffee 1993; Lilly et al. 
1996). In this paper the population of {\it IRAS} galaxies is assumed to undergo 
this form of evolution out to $z=2$, and then to retain its enhanced luminosity 
to a cutoff redshift $z=5$. The same form of evolution was used in models~3, A 
and I2 in Blain \& Longair (1996), Blain (1997a) and Blain (1997b) respectively.
In observations carried out at a wavelength of 850\,$\mu$m after this paper was 
submitted, Smail, Ivison \& Blain (1997) confirmed that very strong evolution of 
the population of distant galaxies is taking place in the submillimetre waveband. 
A surface density of galaxies about 5 times larger than that predicted using the
above model appears to be most consistent with their preliminary conclusions.

\begin{figure}
\begin{center}
\epsfig{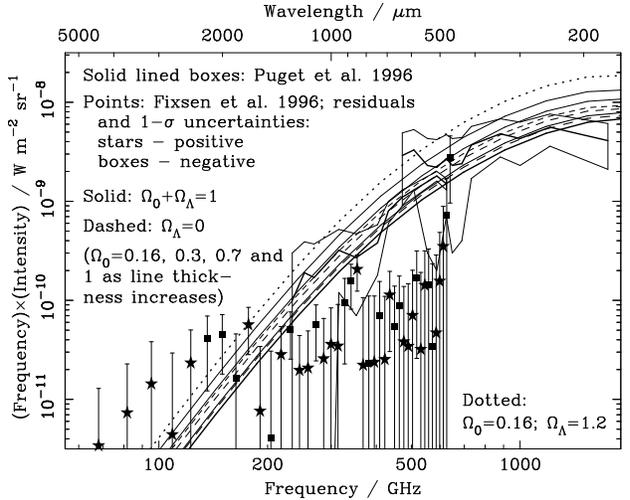}
\end{center}
\caption{The intensity of diffuse background radiation predicted by the model of
galaxy evolution employed here (Section~2.2) in each of the eight world models 
listed in Table~1 (smooth curves). The background intensities and limits derived 
from observations using the FIRAS instrument on the {\it COBE} satellite 
(Fixsen et al. 1994) by Puget et al. (1996) and Fixsen et al. (1996) are represented 
by solid-edged boxes and points with error bars respectively. The residuals
(Table\,4; Fixsen et al. 1996) describe the signal that remains after models 
of both the cosmic microwave background radiation intensity and galactic 
emission are subtracted from the all-sky FIRAS data.}
\end{figure}

\subsection{World models}

The Friedmann equation for the expansion rate takes the form,
\begin{equation}
\dot R^2 = H_0^2 \left[ { {\Omega_0} \over {R} } + 
\Omega_\Lambda R^2 \right] - A^2 c^2, 
\end{equation}
in the notation used here. The scale factor is represented by $R$, the curvature 
parameter $A = (H_0/c) \sqrt{\Omega_0 + \Omega_\Lambda - 1}$ and the density 
parameter of the universe at the present epoch is represented by $\Omega_0$. In 
a flat world model $A=0$ and $\Omega_0 + \Omega_\Lambda = 1$. The world 
models used here are listed in Table~1; four flat models, three open models with 
$\Omega_\Lambda = 0$, 
and one closed model with a large value of $\Omega_\Lambda$.

\begin{figure*}
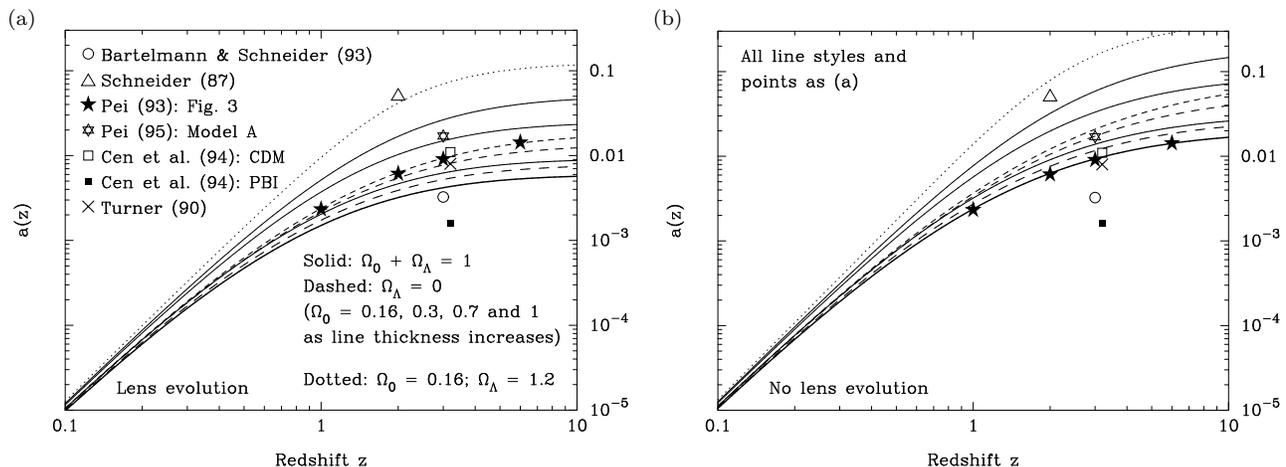

\begin{minipage}{170mm}
(a) \hskip 81mm (b)
\begin{center}
\vskip -5mm
\epsfig{file=FIG2A.PS,width=5.8cm,angle=-90} \hskip 4mm
\epsfig{file=FIG2B.PS,width=5.8cm,angle=-90}
\end{center}
\caption{The redshift dependence of the probability of galaxy--galaxy lensing
predicted in the eight world models listed in Table~1, with (a) and without (b) 
Press--Schechter evolution of the mass distribution of lensing objects. The 
curves are normalised to match the prediction of Pei (1993). The probability of 
lensing in an open model with a density parameter $\Omega_0<1$ and 
$\Omega_\Lambda = 0$ is very similar to that in a flat model with a density 
parameter $(\Omega_0 + 1)/2$ and $\Omega_\Lambda \ne 0$ (Kochanek 1996).}
\end{minipage} 
\end{figure*}

\subsection{Comparison with observations}

The intensity of diffuse extragalactic background radiation and the abundance of 
heavy elements at the present epoch, $\Omega_{\rm m}$ expected in each of the 
world models listed in Table~1 are compared in Fig.~1 and the final column of 
Table~1 respectively; for details of the calculations see Blain \& Longair 
(1993a,b). The abundances were calculated assuming that all of the energy 
emitted by galaxies appears in the far-infrared waveband and is generated by the 
transmutation of hydrogen into heavy elements in massive stars. The efficiency 
of the conversion of rest mass of consumed hydrogen into energy in this process 
is assumed to be 0.7\%. These calculations should yield a conservatively large 
estimate of $\Omega_{\rm m}$, as the contributions made to dust heating, and 
hence to the far-infrared luminosities of galaxies, by both active galactic nuclei 
and non-helium-burning stars are not considered. In a recent review of the 
properties of luminous infrared galaxies Sanders \& Mirabel (1996) reported that 
about 15\% contain active nuclei, while Calzetti et al. (1995) noted that about 
30\% of the luminosity of star-forming galaxies in the far-infrared waveband can 
be generated by the non-ionizing radiation from lower-mass stars. 

The predicted background radiation intensities in Fig.~1 are all broadly 
consistent with the results of Puget et al. (1996), although a generally larger flux 
density is predicted in the closed model. If the density parameter in baryons at 
the present epoch is about 0.05 (Bristow \& Phillipps 1994), then the values of
$\Omega_{\rm m}$ listed in Table~1 correspond to between about 0.9 and 2.6\% 
of the mass of baryons being processed in stars by the present epoch. This range 
of values does not seem unreasonable, as the mass of carbon and heavier 
elements in the sun is approximately 2.4\% of the mass of hydrogen 
(Savage \& Sembach 1996). Hence, the predictions of neither background 
intensities nor metal abundances can be used to discriminate strongly between 
the different world models.

\section{Galaxy--galaxy lensing}

\subsection{Introduction and formalism}

Predictions of the counts of galaxy--galaxy gravitational lenses in the 
submillimetre waveband were made by Blain (1996b) in an Einstein--de Sitter 
world model, based on existing formalism (Peacock 1982; Pei 1995). More 
recently, world models including $\Omega_\Lambda \ne 0$ (Fukugita et al. 1992) 
have been considered (Blain 1997a); however, apart from noting that the 
surface density of lensed galaxies is expected to increase in such a world model, 
the properties of the resulting population of lensed galaxies were not discussed 
in any detail. 

The probability that a galaxy at redshift $z$ is lensed into multiple images by a 
galaxy at a smaller redshift and magnified by a factor between $A$ and 
$A+{\rm d}A$ is $a(z) A^{-3}\,{\rm d}A$. The function $a$ incorporates 
all the details of the world model, which determines the relative distances 
between observer, lens and source, any evolution of the mass distribution of
the population of lensing galaxies and the density parameter in compact lensing 
objects at the present epoch $\Omega_{\rm L}$. The lensing optical depth 
$\tau \simeq a/8$. The mass distribution of lensing galaxies is assumed either 
to remain fixed with the form it takes at the present epoch or to evolve according 
to the prescription of the Press--Schechter formalism (Press \& Schechter 1974), 
in which the mass distribution of collapsed objects is derived by assuming that 
initial gaussian density fluctuations evolve according to the theory of linear 
perturbation growth in an expanding universe (Peebles 1993). The comoving 
space density and mean mass of lensing galaxies are expected to increase and
decrease respectively with increasing redshift in the Press--Schechter model, 
and so the probability of lensing is predicted to be smaller as compared 
with that derived in a non-evolving model (Blain 1996b). 

Estimates of the form of $a$ in each of the world models listed in Table~1 are 
presented in Fig.~2(a) \& (b), assuming evolving and non-evolving mass
distributions respectively. $a$ is normalised to match the predictions of Pei 
(1993), which were derived in an Einstein--de Sitter world model with a 
non-evolving mass distribution. These predictions are similar to those made
by a range of different authors. The predictions for each world model are smaller
in the evolving model. The redshift dependence of $a$ is determined both by the 
values of $\Omega_0$ and $\Omega_\Lambda$ and by the form of evolution of 
the mass distribution, while its absolute normalisation is determined by the 
value of $\Omega_{\rm L}$. The prediction plotted in Fig.~2 due to Pei (1995),
for which the lensing normalisation is about a factor of 2 larger than that
assumed here, corresponds to a density $\Omega_{\rm L} = 0.16$ in lensing 
objects, distributed as $\Omega_{\rm L} = 0.01$, 0.05 and 0.10 in compact
objects, galaxies and clusters respectively. The galaxies and clusters are 
modeled as singular isothermal spheres. Recent work by Maller, Flores \& 
Primack (1997) indicates that the probability of galaxy--galaxy lensing could be 
larger than that predicted by a model of singular isothermal spheres if the 
contribution of a disk component in galaxies is included. Hence, the probability of 
lensing assumed here corresponds to a density of singular isothermal spheres 
$\Omega_{\rm L} \sim 0.08$. If this is an overestimate, then the numbers of 
strongly lensed images predicted in this section will be too large, but probably 
not by a factor greater than about 2.

\begin{figure}
\begin{center}
\epsfig{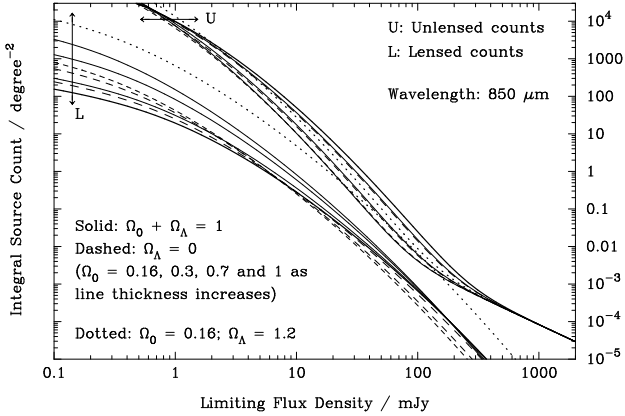}
\end{center}
\caption{The counts of lensed and unlensed galaxies expected in all eight world 
models listed in Table~1. The lensed counts are calculated assuming
Press--Schecter evolution of the populations of lensing galaxies, as 
assumed in Fig.~2(a).}
\end{figure} 

\subsection{Predicted counts of lensed galaxies} 

The counts of lensed galaxies can be predicted by combining the probabilities of 
galaxy--galaxy lensing calculated above with models of the population of 
unlensed background galaxies (Blain 1996b). The counts expected in each of the 
world models listed in Table~1 are compared in Fig.~3, assuming an evolving 
population of lensing galaxies; in Fig.~4 the corresponding ratios between the 
counts of lensed and unlensed galaxies are compared, both with and without
assuming an evolving population of lensing galaxies. 

The magnitude of the ratio of counts of lensed and unlensed galaxies is expected 
to be modified significantly by changing the values of $\Omega_0$ and 
$\Omega_\Lambda$, particularly by increasing the value of $\Omega_\Lambda$, 
and by including evolution of the lensing galaxies. However, the shapes of all 16 
curves presented in Figs~4(a) \& (b) are very similar. The ratio of the counts of 
lensed and unlensed galaxies always has a maximum at the flux density at which 
the unlensed counts steepen significantly as compared with the counts expected 
in a Euclidean model (Blain 1996b). The predicted effect on the counts of 
increasing $\Omega_\Lambda$ is similar to that expected due to either reducing 
the strength of evolution of lensing galaxies or increasing the value of 
$\Omega_{\rm L}$. Hence, if a statistical sample of lensed galaxies can be
compiled in the submillimetre waveband, their counts compared with unlensed 
galaxies can be determined, then a joint limit to the values of $\Omega_{\rm L}$, 
$\Omega_0$ and $\Omega_\Lambda$ and to the evolution of the mass 
function of lensing galaxies can be imposed. 

\begin{figure*}
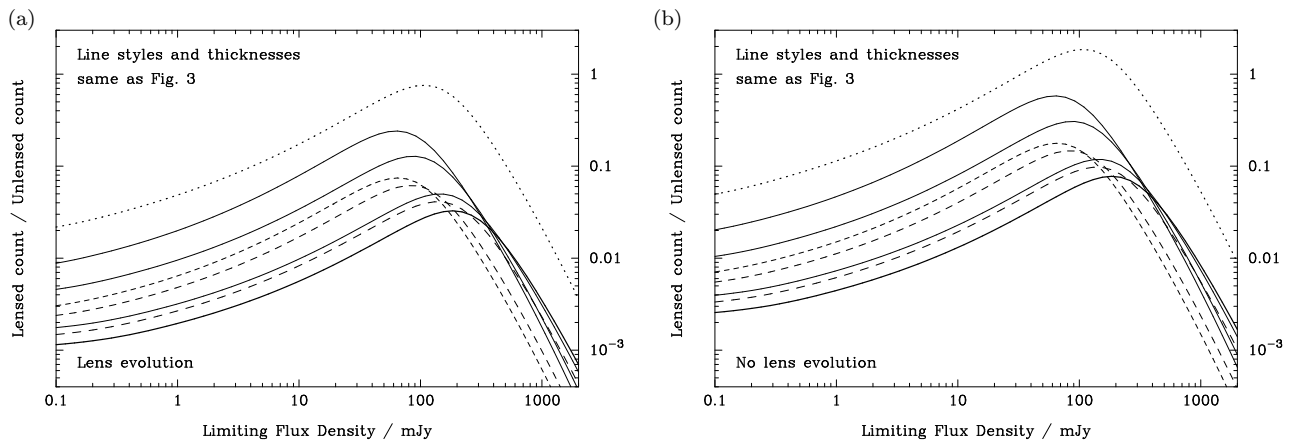

\begin{minipage}{170mm}
(a) \hskip 81mm (b)
\begin{center}
\vskip -5mm
\epsfig{file=FIG4A.PS, width=5.45cm, angle=-90} \hskip 5mm
\epsfig{file=FIG4B.PS, width=5.45cm, angle=-90}
\end{center}
\caption{The relative counts of lensed and unlensed galaxies predicted with 
(a) and without (b) evolution of the population of lensing galaxies. A larger
number of lensed images are expected in the no-evolution model. The curves in 
(a) can be obtained by directly dividing the lensed and unlensed counts
presented in Fig.~3.}
\end{minipage}
\end{figure*} 

Additional information would be required to distinguish between the effects of
these factors. The shape of the unlensed counts could be used to provide
information about both the form of evolution of distant galaxies (Blain \& Longair
1996) and the world model (Fig.~3). The redshifts and detailed image structures 
of individual lenses, derived from follow-up observations in the near-infrared, 
optical and submillimetre wavebands, could be combined with lens modeling 
(for example Williams \& Lewis 1996) to impose additional independent limits to 
cosmological parameters, subject to knowledge of the mass distribution within 
the lensing galaxies.

\subsection{Observing submillimetre-wave lenses} 

Considerable progress is being made in observational cosmology in the 
submillimetre waveband. The counts of galaxies in the submillimetre waveband 
with 850-$\mu$m flux densities of several mJy will soon be determined for the 
first time in a blank-field survey using the new SCUBA bolometer array detector 
(Cunningham et al.~1994) at the James Clerk Maxwell Telescope (JCMT), which 
operates at wavelengths of 450 and 850\,$\mu$m (Blain \& Longair 1996). Smail, 
Ivison \& Blain (1997) have recently determined an estimate of these counts in 
the fields of lensing clusters. However, because the detection of order $10^2$ 
galaxies is expected in a long-term blank-field SCUBA survey, and the fraction
of lenses at a flux density of a few mJy is not expected to be larger than about 
$10^{-2}$, it is unlikely that more than a handful of lensed galaxies will be 
detected in a deep blank-field SCUBA survey; the 6-arcsec angular resolution of 
the JCMT will also not resolve the different components of multiply-imaged 
lenses (Blain 1996b).

The next generation of submillimetre-wave telescopes -- MIAs, other 
ground-based telescopes, {\it FIRST} and {\it Planck Surveyor} -- will make the 
detection and investigation of lensed galaxies a key part of submillimetre-wave 
astronomy. A more detailed discussion of the strategy for detection and 
identification of lensed galaxies in future submillimetre-wave surveys is provided 
by Blain (1997a,c). First, the strategy relies on the detection of galaxies with flux 
densities corresponding to the peak of the curves in Fig.\,4 in large-area surveys 
using the {\it Planck Surveyor} and {\it FIRST} telescopes. This ensures that the 
fraction of lensed galaxies in the total sample is as large as possible. Secondly,  
a large MIA would be used for high-resolution follow-up observations to identify 
the galaxies that show signs of arcs and multiple images. Thirdly, the
MIA-selected candidates would also be observed in the near-infrared, optical and 
radio wavebands in order to determine the detailed properties of the lens and 
source. 

In a large-area survey using the {\it FIRST} telescope, and an all-sky survey 
using {\it Planck Surveyor}, catalogues containing of order $10^4$ and $10^5$ 
galaxies and AGN with flux densities greater than about ten and several tens of 
mJy respectively are expected (Blain 1997a,c). These large catalogues will yield 
accurate submillimetre-wave source counts. The catalogues will not be affected 
significantly by source confusion (Blain, Ivison \& Smail 1997) -- a typical 
separation of about 10\,arcmin is expected for sources brighter than 10\,mJy
(Fig.\,3). However, because the resolution of these telescopes is coarse as 
compared with ground-based submillimetre-wave telescopes, the sources will 
not be resolved. All of the detected galaxies, which are selected solely because 
of their large flux densities in the submillimetre waveband, are candidates for 
further investigation as potential lenses; however, an optimized selection could 
be made if submillimetre/far-infrared colours were determining for these galaxies 
in a {\it FIRST} survey (Blain 1997a).

The sub-arcsecond angular resolution and excellent sensitivity of an MIA 
will allow the rapid resolution and identification of any sources containing 
structures that resemble arcs or multiple images in the {\it FIRST} and 
{\it Planck Surveyor} catalogues. Most catalogued sources are expected to 
be distant active or vigorous star-forming galaxies, and so are unlikely to be 
extended or to contain significant substructure that could be misinterpreted as 
a signature of lensing. The far-infrared luminosity of merging luminous starburst 
galaxies at small redshifts is dominated by the emission from a single compact
core region, which is typically only several hundred parsecs across (Solomon 
et~al. 1997). Hence, the detection of arcs or multiple images in an MIA follow-up 
observation would provide a strong indication that a lensed galaxy had been 
found. Based on the size of the {\it FIRST} and {\it Planck Surveyor} catalogues 
and the estimated fractions of lensed galaxies shown in Fig.\,4, a catalogue
containing several hundreds of galaxy--galaxy lenses across a large area
of sky could be compiled.

The selection method ensures that the flux densities of the candidates at a 
wavelength of 850\,$\mu$m will be at least 10\,mJy, corresponding to a 
bolometric luminosity of order 10$^{13}$\,L$_\odot$. Hence, follow-up
observations of these very luminous sources in other wavebands should be 
relatively easy. The candidates are expected to have properties similar to 
that of the archetypal lensed ultraluminous galaxy {\it IRAS} F10214+4724 
(Rowan-Robinson et al. 1991), which has a B-band magnitude of about 22. 
Lensed structures were clearly detected in near-infrared images of F10214 using
a 4\,m-class telescope in a 90-minute integration (Close et al. 1995). The flux 
densities of the candidate sources are expected to be sufficiently large for many
of them to appear in the {\it IRAS} faint source catalogue; however, they will 
constitute only a very small proportion of the total number of {\it IRAS} sources.

The accurate counts of lensed and unlensed galaxies derived from this 
programme of observations could be compared with model predictions, such as 
those shown in Figs\,3 and 4, in order to investigate the form of the world model 
and the growth of cosmic structure. As discussed by Blain (1997a,c), this
programme would require a large, but not impractical, amount of observing time; 
several months of dedicated observations with an MIA in order to identify lensed
structures, and many nights of spectroscopic observations using large 
telescopes in the near-infrared and optical wavebands in order to determine the 
redshifts of the detected sources and lensing galaxies.  

\begin{figure*}
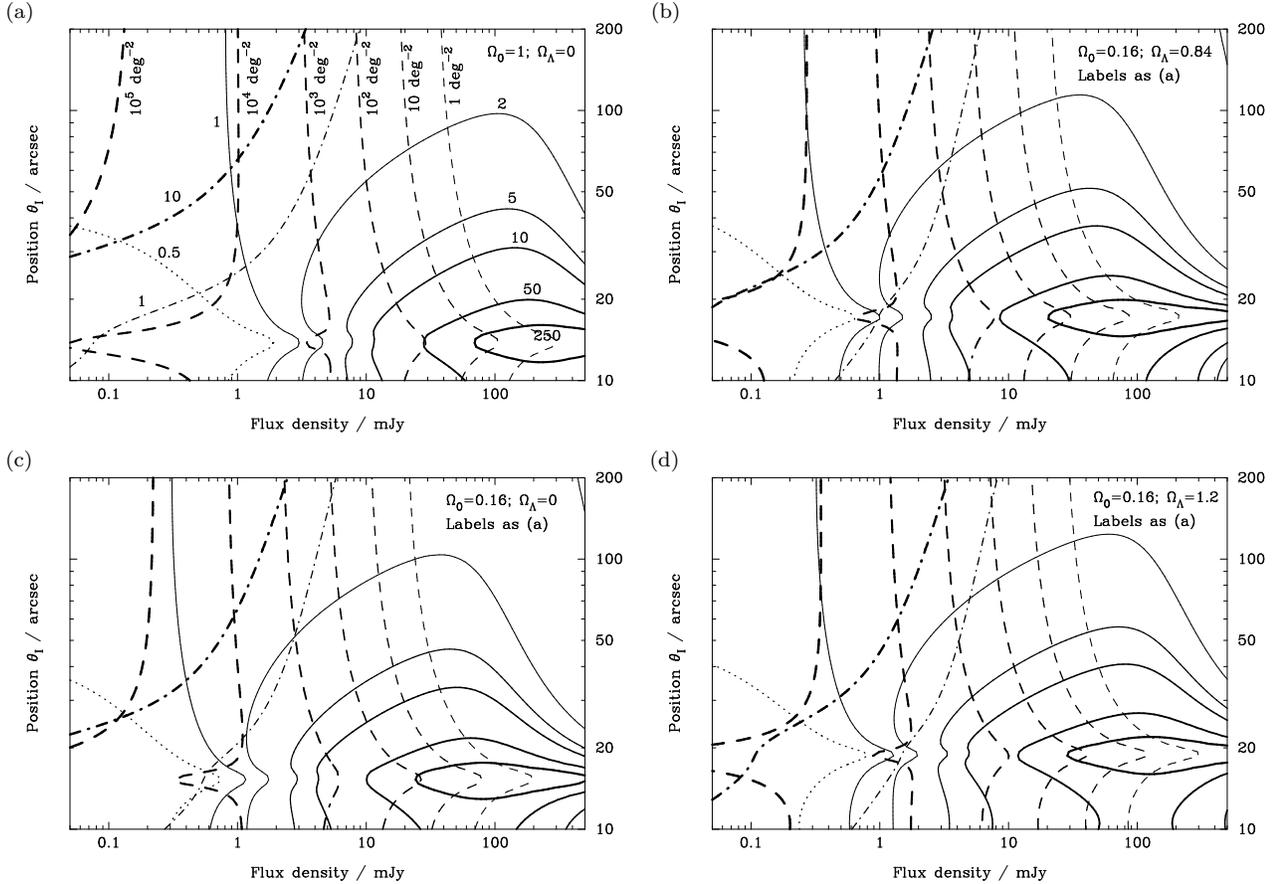

\begin{minipage}{170mm}
(a) \hskip 81mm (b)
\begin{center}
\vskip -5mm
\epsfig{file=FIG5A.PS, width=5.45cm, angle=-90} \hskip 5mm
\epsfig{file=FIG5B.PS, width=5.45cm, angle=-90}
\end{center}
(c) \hskip 81mm (d)
\begin{center}
\vskip -5mm
\epsfig{file=FIG5C.PS, width=5.45cm, angle=-90} \hskip 5mm
\epsfig{file=FIG5D.PS, width=5.45cm, angle=-90}
\end{center}
\caption{The expected lensing properties of a cluster similar to Abell~2218 at 
$z=0.171$ as a function of the flux density of the images and their angular 
distances from the core of a lensing cluster $\theta_{\rm I}$. A spherical
approximation to the mass distribution of Abell~2218 is used (Natarajan \& Kneib 
1996); this reproduces the total mass and approximate density profile of the
cluster. The counts of lensed images of distant dusty galaxies (dashed 
lines), the magnification bias (solid lines for positive bias; dotted lines for 
negative bias) and the number of images expected within a radius 
$\theta_{\rm I}$ (dot-dashed lines) are compared in four different world models
(Table~1).
}
\end{minipage}
\end{figure*} 

\section{Lensing by clusters} 

The effects of lensing by a cluster on the population of background galaxies can 
be considered in terms of a magnification bias, which modifies the counts of
lensed images as compared with background galaxies (Borgeest et al. 1991; 
Broadhurst, Taylor \& Peacock 1995). The bias is introduced because both the 
flux densities and mean separations of the images of background galaxies are 
increased by a lens, and its magnitude is determined by the steepness of the 
slope of the counts of background galaxies. In general these counts are expected 
to be steeper in the submillimetre waveband as compared with the optical 
waveband (Blain 1997b), and so the bias is expected to be relatively large in the 
submillimetre waveband. The angular size of the region within which the bias is 
expected to be significant depends on the relative size of the observer--lens 
and lens--source distance. The forms of both the counts of background galaxies 
(Fig.~3; Blain 1997b) and these distances are expected to depend on the form of 
the world model, and so detailed observations of the properties of lensed images 
in clusters could be used to constrain the values of cosmological parameters. 

The predicted surface density and magnification bias of lensed images in a rich 
cluster are compared in Fig.~5 as a function of flux density and position in the
cluster for four world models selected from the list in Table~1. The expected 
number of detectable images within each radius are also shown. The details of
the calculations are described in Blain (1997b). The form of the magnification bias 
is predicted to differ considerably between world models, particularly at flux 
densities of about 100\,mJy and radii of about 30\,arcsec, for which the bias factor 
is expected to exceed 100. However, the surface density of images is not 
expected to be sufficiently large for any images to be detected in this region, as 
shown by the dot-dashed contours in Fig.~5. Only at fainter flux densities of 
between about 0.1 and 1\,mJy, for which the magnification biases are predicted to 
be less dramatic, would a reasonable number of detectable images be expected. 
Note that Smail, Ivison \& Blain (1997) detected six sources within 1.1\,arcmin of
the core of two clusters to a 850-$\mu$m flux density limit of about 2\,mJy using
a 30-hour integration at the JCMT. 

A large MIA should be able to map a field 1\,arcmin in radius at a 5-$\sigma$ flux 
density limit of 0.1\,mJy at a wavelength of 850\,$\mu$m in about 5\,hours (Brown 
1996). The number of images expected to be detected in such an observation, 
with and without the effects of lensing, are listed in Table~2 for all the world 
models listed in Table~1. Hence, sensitive, but practical, submillimetre-wave 
observations of clusters could probably be used to impose limits to the values of 
cosmological parameters. Note, however, that the properties of the population of 
lensed images are expected to depend more sensitively on the form of galaxy 
evolution, as discussed by Blain (1997b), as compared with the world model.
Their properties are also expected to depend on the lensing behaviour of the 
cluster (Natarajan \& Kneib 1996). Hence, the world model could only be 
investigated in detail after both the form of evolution of distant galaxies, and the 
lensing properties of the cluster of interest had been determined accurately. 

\begin{table}
\caption{The numbers of lensed images $N$ expected with flux densities larger
than 0.1 and 1\,mJy at 850\,$\mu$m within 60\,arcsec of the core of the model 
lensing cluster discussed in Section~4. A large MIA could survey this area of sky 
to a 5-$\sigma$ sensitivity of 0.1\,mJy in about 5\,hours (Blain 1997b; Brown 
1996).} 
\begin{tabular}
{ p{0.6cm} p{0.6cm} p{1.4cm} p{1.9cm} p{1.8cm} } 
$\Omega_0$ & $\Omega_\Lambda$ & Geometry & $N(>0.1\,{\rm mJy})$ & 
$N(>1\,{\rm mJy})$ \\
 & &  & Lensed (Unlensed) & Lensed (Unlensed) \\
\end{tabular}
\begin{tabular}
{ p{0.6cm} p{0.6cm} p{1.4cm} p{0.55cm} p{1.0cm} p{0.4cm} p{0.75cm} }
\noalign{\vskip 2mm}
1.0 & 0.0 & Flat & 60 & (110) & 7 & (7) \\
0.7 & 0.3 & Flat & 76 & (140) & 7 & (6) \\
0.3 & 0.7 & Flat & 130 & (220) & 8 & (5) \\
0.16 & 0.84 & Flat & 180 & (290) & 8 & (5) \\
0.7 & 0.0 & Open & 74 & (130) & 7 & (6) \\
0.3 & 0.0 & Open & 110 & (180) & 7 & (5) \\
0.16 & 0.0 & Open & 130 & (200) & 7 & (4)\\
0.16 & 1.2 & Closed & 230 & (400) & 14 & (8) \\
\end{tabular}
\end{table}

\section{Summary}

The fraction of detectable gravitationally-lensed galaxies is expected to be 
significantly larger in the submillimetre waveband as compared with the optical 
and radio wavebands. Detecting lenses will be an important goal for the next 
generation of submillimetre-wave telescopes. Both ground-based interferometer 
arrays (MIAs) and the space-borne telescopes {\it FIRST} and {\it Planck 
Surveyor} will be able to detect large samples of galaxy--galaxy lenses, in both 
large-area (Blain 1997a) and very sensitive small-area (Blain 1996b) surveys.  
An MIA can also observe lensing by clusters in great detail (Blain 1997b). An 
analysis of the properties of a large sample of lenses selected in the 
submillimetre waveband could be used to investigate the form of evolution of 
distant dusty star-forming galaxies, the values of cosmological parameters, the 
total density in lensing objects and the form of evolution of structure in the 
Universe.
\begin{enumerate}
\item Several hundred lensed galaxies and several hundred thousand unlensed 
galaxies are expected in a whole-sky survey using the {\it Planck Surveyor} 
satellite, a large-area survey using {\it FIRST} and surveys of smaller fields using 
a MIA, providing that the population of distant dusty galaxies evolves in a similar 
manner to that of quasars. The relative counts of lensed and unlensed galaxies 
can be used to impose joint constraints on the geometry of the universe, the 
total mass of lensing galaxies and the evolution of cosmic structure. 
\item The effects of each of these factors could be distinguished using
follow-up observations of a subsample of these lenses, using both an MIA to 
resolve their spatial structure and telescopes operating in optical and 
near-infrared waveband to determine the redshifts of both the lens and images. 
Detailed lens modeling, as demonstrated for observations in the optical and radio 
wavebands, could then be used to investigate the world model alone.  
\item 
The surface and flux density distributions of lensed images in the fields of 
clusters are expected to depend on both the intrinsic properties of the 
background galaxies and the form of the world model. Observations using an 
MIA are potentially very useful for detecting these images. If the form of 
evolution of the population of distant star-forming galaxies is determined 
accurately in blank-field submillimetre-wave surveys then these observations 
can be used to impose direct constraints on the values of cosmological 
parameters.
\end{enumerate}

\section*{Acknowledgements}
I would like to thank Malcolm Longair for his helpful comments on the 
manuscript, and an anonymous referee for his/her very prompt and helpful 
comments.

\end{document}